\documentclass[prb,12pt]{revtex4}
\usepackage[english]{babel}
\usepackage[intlimits]{amsmath}
\usepackage[dvips]{graphicx}
\usepackage{epsfig}

\begin{document}

\begin{titlepage}

\title{Effect of Interactions on Molecular Fluxes and Fluctuations in the Transport Across Membrane Channels}
\author{Anatoly B. Kolomeisky and Stanislav Kotsev}
\affiliation{Department of Chemistry, Rice University, Houston, TX 77005-1892}

\begin{abstract}
Transport of molecules across  membrane channels is investigated theoretically using exactly solvable one-dimensional discrete-state stochastic models. An interaction between  molecules and membrane pores is modeled via a set of binding sites  with different energies. It is shown that the interaction potential strongly influences the particle currents as well as fluctuations in the number of translocated molecules. For small concentration gradients the attractive sites lead to largest currents and fluctuations, while  the repulsive interactions yield the largest fluxes and dispersions for large concentration gradients. Interaction energies that lead to maximal currents and maximal fluctuations are the same only for locally symmetric potentials, while they differ for  the locally asymmetric potentials. The conditions for the most optimal translocation transport with maximal current and minimal dispersion are discussed. It is argued that in this case the interaction strength is independent of local symmetry of the potential of mean forces. In addition, the effect of the global asymmetry of the interaction potential is investigated, and it is shown that it also strongly affects the particle translocation dynamics. These phenomena can be explained by analyzing the details of the particle entering and leaving the binding sites in the channel.

\end{abstract}

\maketitle

\end{titlepage}

\section{Introduction}

Membrane channels are large water-filled hollow protein structures  that control the transport of metabolite molecules between different cells or between different cellular compartments.\cite{lodish_book,hille_book}  These processes are critically important for biological systems, and recent experimental and computational studies suggest that, contrary to earlier views,  permeation of molecules across such large channels is efficient and  selective.\cite{rostovtseva98,hilty01,kullman02,nestorovich02,schwarz03,lu03,danelon06}   However, our understanding of these phenomena is still very limited. 

To analyze mechanisms of efficiency and selectivity of the transport across membrane channels several theoretical methods have been presented.\cite{chou,nelson02,berezhkovskii02,berezhkovskii03,berezhkovskii05,bezrukov07,kolomeisky07} One approach utilizes a continuum description where a single-molecule transport across membrane pores is viewed as an effective one-dimensional diffusion in a potential of mean forces created by interactions between the solute and protein channel.\cite{berezhkovskii02,berezhkovskii03,berezhkovskii05,bezrukov07} The interactions are modeled as square well potentials that occupy the entire channel. Using this method, it was shown  that the particle's current can be maximized for some interaction strength that depends on the solute concentrations, diffusion constants and geometry of the pore.\cite{berezhkovskii05} It is also  possible to compute the inter-channel potential of mean force that maximizes the flux.\cite{bezrukov07}  We recently developed a discrete-state stochastic model of the channel-facilitated membrane transport that takes into account the existence of binding sites inside the pore.\cite{kolomeisky07} By mapping the discrete-state model of the permeation through the pore to a single-particle hopping along a periodic lattice, the particle currents have been obtained explicitly for all sets of parameters. The theoretical analysis suggested that the presence of the binding sites accelerates the particle flux for small concentration gradients, while the repulsive binding sites are more advantageous for creating the most optimal current for large concentration differences. In addition, it was shown that the asymmetry in the interaction potential, e.g., the spatial position of the binding sites, might also significantly change the particle dynamics. Similar observations have also been  obtained in the continuum models of membrane transport.\cite{berezhkovskii05,bezrukov07}  Theoretical calculations show that both continuum and discrete-state descriptions are closely related,\cite{berezhkovskii05,bezrukov07} and the results obtained by these approaches can be mapped into each other.

In many biological systems  concentrations of molecules that involved in permeation trough the membrane pores are rather small, and this points out to the importance of fluctuations in the number of translocated particles.\cite{hille_book} However, current theoretical studies concentrate mostly on the description of fluxes, i.e., the average number of particles moved across a single pore  per unit time.\cite{chou,nelson02,berezhkovskii02,berezhkovskii03,berezhkovskii05,bezrukov07,kolomeisky07}  In this work, we analyze the effect of interactions on fluctuations in the number of translocating molecules and compare it with the effect of interaction potentials on the particle currents. Our approach is based on the discrete-state stochastic models for which {\it all} dynamic properties can be calculated  explicitly via mapping to the single-particle random hopping model on  periodic lattices.\cite{kolomeisky07,AR07}

\section{Model}

We consider a transport of particles through a membrane channel  as an effective one-dimensional motion across a cylindrical pore with $N$ binding sites inside. The pore  separates two chambers as illustrated in Fig. 1. The molecule's concentrations in the left and right chambers are $c_{1}$ and $c_{2}$, respectively. The concentration gradient  $\Delta c=c_{1}-c_{2}$ ($c_{1}>c_{2}$) drives the particle current  mostly from the left to the right.  The molecule can enter  the channel from the left  with the rate $u_{0}=k_{on} c_{1}$, but from the first binding site it can also return back with the rate $w_{1}=k_{off}$. Similarly, the molecule can jump to the pore from the right chamber or jump out from the state $N$ with  rates $w_{N+1}=w_{0}=k_{on} c_{2}$ and $u_{N}=k_{off}$, correspondingly. The particle at the site $j$ ($j=1,2,\cdots,N$) can move one site forward with the rate $u_{j}$, or it can jump backward one site with the rate $w_{j}$: see Fig. 1. Because the concentration of solute molecules is typically small, it is assumed that particles do not interact with each other, and there is no more than one particle can be found in the channel at all times. The probability to find the particle at the binding site $j$ at time $t$ is given by a function $P_{j}(t)$, and time evolution of the translocation process can be described by set of master equations, 
\begin{equation}\label{master}
\frac{dP_{j}(t)}{dt}=u_{j-1}P_{j-1}(t)+w_{j+1}P_{j+1}(t)-(u_{j}+w_{j}) P_{j}(t),
\end{equation}
where $j=1,2,\cdots,N$, and we have defined
\begin{equation}
P_{0}(t) \equiv P_{N+1}(t)=1 - \sum_{j=1}^{N} P_{j}(t)
\end{equation}
as the probability of finding the channel empty and the  molecule outside of the pore at time $t$.

Each binding site corresponds to a minimum in the free energy profile for translocation as illustrated in Fig. 2. We associate the strength of the interaction at site $j$ with the parameter $(-\varepsilon_{j})$, assuming that zero free energy is at the entrance, $\varepsilon_{0}=0$.\cite{kolomeisky07} Note that the free energy at the exit to the right chamber is $(-\varepsilon_{N+1})$. Then, $\varepsilon_{j}>0$ describe the attractive binding sites (with respect to the left chamber), while negative $\varepsilon_{j}$ corresponds to the repulsive sites.\cite{kolomeisky07} The transition rates between the  sites are related to binding energies via detailed balance conditions, 
\begin{equation}\label{detailed_balance}
\frac{u_{j}(\varepsilon_{j},\varepsilon_{j+1})}{w_{j+1}(\varepsilon_{j},\varepsilon_{j+1})}=\frac{u_{j}(\varepsilon_{j},\varepsilon_{j+1}=0)}{w_{j+1}(\varepsilon_{j},\varepsilon_{j+1}=0)} x_{j+1}=\frac{u_{j}(\varepsilon_{j}=0,\varepsilon_{j+1})}{w_{j+1}(\varepsilon_{j}=0,\varepsilon_{j+1})}(1/ x_{j}),
\end{equation}
with $x_{j}=\exp{[\varepsilon_{j}/k_{B}T]}$. Dynamic properties of the system depend on explicit expressions for the transition rates that can be written in the following form,\cite{kolomeisky07}
\begin{equation}
u_{j}(\varepsilon_{j+1})=u_{j}(\varepsilon_{j+1}=0)x_{j}^{\theta_{j}}, \quad w_{j+1}(\varepsilon_{j+1})=w_{j+1}(\varepsilon_{j+1}=0)x_{j}^{\theta_{j}-1},
\end{equation}
or 
\begin{equation}
u_{j}(\varepsilon_{j})=u_{j}(\varepsilon_{j}=0)x_{j}^{\theta_{j}-1}, \quad w_{j+1}(\varepsilon_{j})=w_{j+1}(\varepsilon_{j}=0)x_{j}^{\theta_{j}},
\end{equation}
where $0 \le \theta_{j} \le 1$ are interaction-distribution coefficients that describe how binding energies are distributed between forward and backward transition states. These coefficients also provide relative distances between neighboring free-energy minima and transition states (maxima in Fig. 2).  Similar parameters have been utilized in the analysis of motor protein's dynamics.\cite{AR07} To simplify calculations, in this paper we will assume that interaction-distribution coefficients are the same for all binding sites, i.e., $\theta_{j}=\theta$ for all $j$.

If we define $M(t)$ as number of particles that translocated through the membrane channel at time $t$, then the stationary-state particle flux is given by
\begin{equation}
J=\lim_{t \rightarrow \infty} \frac{d \langle M(t) \rangle}{d t},
\end{equation}
where averaging is taken over all possible translocation events. To specify fluctuations, a dispersion $D$ is introduced in the following way,
\begin{equation}
D=\frac{1}{2}\lim_{t \rightarrow \infty} \frac{d \left( \langle M^{2}(t) \rangle -\langle M(t) \rangle^{2} \right)}{d t}.
\end{equation}

The discrete-state stochastic model of channel-facilitated membrane transport can be solved exactly at large times by utilizing the mapping to the single-particle hopping model along infinite one-dimensional periodic chain.\cite{kolomeisky07} This mapping can be understood using the following arguments. Consider multiple identical membrane channels (as shown in Fig. 1) arranged in a sequence such that the particle exited from one channel can enter the next one. At the stationary state the flux is constant, and the transport of the particles along the sequence of channels (with $N$ binding sites in each) is identical to the motion of the single particle along one-dimensional periodic lattice with a period of $N+1$ sites. The number of states in the period of the effective lattice is larger than in the channel because the additional state corresponds to the situation when the particle is outside of the channel. For the effective single-particle hopping model {\it all} dynamic properties are known exactly.\cite{derrida83,AR07} Thus, transport across membrane pores can be analyzed explicitly for all sets of parameter in the stationary-state limit.

\section{Results and Discussion}

To investigate the effect of interactions on particle dynamics in channel-facilitated membrane transport we consider the simplest model with $N=1$ binding site inside the pore.\cite{kolomeisky07}  The binding energy is equal to $-\varepsilon$, and the detailed balance conditions can be written as
\begin{equation}\label{db1}
\frac{u_{0}(\varepsilon)}{w_{1}(\varepsilon)}=\frac{u_{0}(\varepsilon=0)}{w_{1}(\varepsilon=0)} x, \quad \frac{u_{1}(\varepsilon)}{w_{0}(\varepsilon)}=\frac{u_{1}(\varepsilon=0)}{w_{0}(\varepsilon=0)} (1/x),
\end{equation}
where $x=\exp{[\varepsilon/k_{B}T]}$. The transition rates are given by\cite{kolomeisky07}
\begin{equation}
u_{0}(\varepsilon)=u_{0} x^{\theta}, \quad w_{1}(\varepsilon)=w_{1} x^{\theta-1},\quad u_{1}(\varepsilon)=u_{0} x^{\theta-1}, \quad w_{0}(\varepsilon)=w_{0} x^{\theta}.
\end{equation}
Then using known results,\cite{derrida83,AR07} the particle flux for $N=1$ model has a simple form,
\begin{equation}\label{flux_n2}
J=\frac{(u_{0}u_{1}-w_{0}w_{1})x^{\theta}}{(u_{0}+w_{0})x+(u_{1}+w_{1})}=\frac{k_{on}(c_{1}-c_{2})x^{\theta}}{2+\frac{k_{on}(c_{1}+c_{2})}{k_{off}}x}.
\end{equation}
The corresponding expression for the dispersion is more complex,\cite{AR07}
\begin{equation}
D=\frac{1}{2}\frac{(u_{0}u_{1}+w_{0}w_{1})x^{\theta}-2\frac{(u_{0}u_{1}-w_{0}w_{1})^{2}x^{\theta+1}}{[(u_{0}+w_{0})x+u_{1}+w_{1}]^{2}}}{(u_{0}+w_{0})x+u_{1}+w_{1}}=\frac{1}{2} \frac{k_{on}k_{off}(c_1+c_2)x^{\theta}-2\frac{k_{on}^2k_{off}^2(c_1-c_2)^2}{[k_{on}(c_1+c_2)x+2 k_{off}]^2}x^{\theta+1}}{k_{on}(c_1+c_2)x+2 k_{off}}.
\end{equation}
It is convenient to consider the relative flux,
\begin{equation}\label{relative_flux_n2}
\frac{J} {J_0}=\left[ \frac{k_{on}(c_1+c_2)+2 k_{off}}{k_{on}(c_1+c_2)x+2 k_{off}} \right] x^{\theta}
\end{equation}
and the relative dispersion, 
\begin{equation}\label{relative_D_n2}
\frac{D} {D_0}=\left[\frac{k_{on}(c_1+c_2)+2 k_{off}}{k_{on}(c_1+c_2)x+2 k_{off}}\right] x^{\theta}
\left( \frac{k_{on}k_{off}(c_1+c_2)-2\frac{k_{on}^2k_{off}^2(c_1-c_2)^2}{[k_{on}(c_1+c_2)x+2 k_{off}]^2}x}{k_{on}k_{off}(c_1+c_2)-2\frac{k_{on}^2k_{off}^2(c_1-c_2)^2}{[k_{on}(c_1+c_2)+2 k_{off}]^2}} \right),
\end{equation}
 where $J_{0}$ and $D_{0}$ are dynamic properties of the system without interactions ($\varepsilon=0$). 

The molecular flux and dispersion are strongly influenced by interactions at the binding site, as illustrated in Fig. 3. The presence of strongly attractive or strongly repulsive binding sites lead to decrease in both particle currents and fluctuations. For intermediate interactions the molecular flux and dispersion are large, and this behavior is independent of concentration gradients. The relative particle current reaches a maximum value at the interaction energy $\varepsilon_{J}^{*}$ that can be obtained from Eq. (\ref{relative_flux_n2}), 
\begin{equation}\label{max_flux_n2}
\varepsilon_{J}^* = k_B T \ln{\left[\frac{\theta}{1-\theta}\ \frac{2 k_{off}}{k_{on}(c_1+c_2)}\right]}.
\end{equation}
Fluctuations in the number of translocating particle also produce a maximum as a function of the binding energy strength, as can be seen in Fig. 3, but the corresponding energy of interactions $\varepsilon_{D}^{*}$ yields a more complex expression, namely,
\begin{eqnarray}\label{max_D_n2}
\varepsilon_{D}^{*}=k_B T \ln{ \left[ G\left(\frac{\theta}{1-\theta}\right)\frac{ k_{off}}{k_{on}c_1} \right]},
\end{eqnarray}
for the simplest case of $c_{2}=0$. The auxiliary function $G(\gamma)$ is defined as
\begin{eqnarray}\label{G}
G(\gamma)&=&\sqrt[3]{\frac{64}{27} \gamma^3 - \frac{4}{3}\gamma + \sqrt{\frac{512\gamma^4 -976\gamma^2 +512}{27}} }\nonumber \\
&+&\sqrt[3]{\frac{64}{27} \gamma^3 - \frac{4}{3 }\gamma - \sqrt{\frac{512\gamma^4 -976\gamma^2 +512}{27}}} + \frac{4}{3} \gamma. 
\end{eqnarray}
It has the following properties that $G(\gamma=1)=2$, and for small $\gamma$ we have $G(\gamma) \simeq \gamma/3$, while for $\gamma \gg 1$ the asymptotic behavior is different, $G(\gamma) \simeq 4 \gamma$.

Interactions energies $\varepsilon_{J}^{*}$  and $\varepsilon_{D}^{*}$ that produce maximal fluxes and maximal dispersions depend on the concentrations outside of the membrane channel and on the local environment around the binding site (via the interaction-distribution parameter $\theta$) as shown in Fig. 4. When the concentration of particles in the left chamber is small the presence of attractive site leads to the largest particle current, and at these conditions fluctuations are also maximal. However, for large concentrations $c_{1}$ the repulsive site produces the biggest flux and dispersion. This behavior has been explained before by considering the details of the particle dynamics near the binding site.\cite{kolomeisky07} It can be shown that the total time to move across the channel consists of two terms that describe the effective time to enter and to leave the binding site.\cite{kolomeisky07} The maximal current is achieved when these two terms are of the same order. Then for small concentration gradients the effective time to enter the binding site is large, and to produce the optimal flux it is required that the particle stays longer in the channel, which corresponds to $\varepsilon_{J}^{*}>0$. For large values of $c_{1}$ the time to move into the binding site is small, and only repulsive interactions will lower the time for the particle in the channel, producing the largest  current. It is reasonable to suggest that similar arguments can be used to explain the behavior of dispersion. 

The results presented in Fig. 4b also show that interaction energies that lead to maximal fluxes and dispersions strongly depend on the location of transition states near the binding site. When the position of the transition state between two minima in the potential of mean forces is closer to the right one  ($\theta >0.5$), the binding interactions that produce maximal fluctuations are larger than the interactions that yield the maximal flux. At the same time, for the transition state closer to the left state ($\theta <0.5$) the situation is different, and $\varepsilon_{J}^{*} > \varepsilon_{D}^{*}$.  Thus for the locally symmetric potential of mean forces ($\theta=0.5$) the interaction energies $\varepsilon_{J}^{*}$  and $\varepsilon_{D}^{*}$ coincide, while the local asymmetry ($\theta \ne 0.5$) yields different values for the most optimal binding sites interactions. This observation can be understood by analyzing Eqs.(\ref{relative_flux_n2}) and (\ref{relative_D_n2}). Comparing these two expressions for the maximal interactions strengths it can be shown that
\begin{equation} \label{Df}
\frac{D} {D_0}=\frac{J}{J_{0}} F(x),
\end{equation}
where the function $F(x)$, the ratio of the relative dispersion over the relative flux, is given by
\begin{equation}\label{Fx}
F(x)=   \frac{k_{on}k_{off}(c_1+c_2)-2\frac{k_{on}^2k_{off}^2(c_1-c_2)^2}{[k_{on}(c_1+c_2)x+2 k_{off}]^2}x}{k_{on}k_{off}(c_1+c_2)-2\frac{k_{on}^2 k_{off}^2(c_1-c_2)^2}{[k_{on}(c_1+c_2)+2 k_{off}]^2}}.
\end{equation}
Taking derivative of the left and right side of Eq. (\ref{Df}) with respect to the variable $x$, we obtain
\begin{eqnarray} \label{derivative}
\frac{d (D/D_0)}{d x} & = &  \nonumber \\
 \frac{d (J/J_{0})}{d x} F(x) & - & 2\frac{J}{J_{0}} \frac{k_{on}^{2} k_{off}^2(c_1-c_2)^{2}}{[k_{on}(c_1+c_2)x+2 k_{off}]^{3}}\frac{2k_{off}-x k_{on}(c_{1}+c_{2})}{k_{on}k_{off}(c_1+c_2)-2\frac{k_{on}^2 k_{off}^2(c_1-c_2)^2}{[k_{on}(c_1+c_2)+2 k_{off}]^2}}.
\end{eqnarray}
Then $\frac{d (D/D_0)} {d x} $ and $\frac{d (J/J_{0})}{d x}$ simultaneously become equal to zero only if
\begin{equation}
x=\frac{2k_{off}}{k_{on}(c_{1}+c_{2})}.
\end{equation}
Comparing this result with Eq.(\ref{max_flux_n2}) and recalling that $x=\exp{(\varepsilon/k_{B}T)}$ it can be concluded that $\varepsilon_{J}^{*}=\varepsilon_{D}^{*}$ only for $\theta=1/2$.

The observation that the maximal particle current and the maximal dispersion for $\theta \ne 0.5$ are realized for different interaction strengths raises the question of what is the  binding interaction energy that allows us to have the largest possible current simultaneously with smallest fluctuations in the number of translocated molecules. From the point of view of functionality of cellular processes this interaction strength might be viewed as the most optimal. To answer this question we analyze the function $F(x)$, the ratio of the relative dispersion over the relative current, as  given in Eq. (\ref{Fx}). The most optimal conditions can be reached by minimizing this function.  It can be shown that this ratio is minimal when
\begin{equation}\label{optimal}
\varepsilon_{opt}^* = k_B T \ln{\left[\frac{2 k_{off}}{k_{on}(c_1+c_2)}\right]}.
\end{equation}
Note that, as discussed above, for the locally symmetric potentials ($\theta=0.5$) we have $\varepsilon_{J}^{*}=\varepsilon_{D}^{*}=\varepsilon_{opt}^*$. However, generally the most optimal interaction does not produce the largest fluxes or fluctuations, and it is independent of the interaction-distribution parameter $\theta$.  These results are illustrated in Fig. 5.

Theoretical and experimental investigations of the potential of mean forces\cite{jensen02,alcaraz04,kosztin04} indicate that the free-energy landscape for the molecules permeating  across membrane channels is generally globally asymmetric. This asymmetry is important for biological channels,\cite{kosztin04,kolomeisky07}  and for  the transport across artificial pores.\cite{shaw07} It was shown recently by one of us\cite{kolomeisky07} that the asymmetry influences the particle fluxes, and the origin of this phenomenon was discussed by analyzing the dynamics of particle translocation.  To study the effect of asymmetry on fluctuations in the number of translocating particle, we consider two membrane channel models with $N=2$ binding sites. In the first model, the binding energies on sites 1 and 2 are equal to $(-\varepsilon)$ and 0, correspondingly, while in the second model the order is reversed. Otherwise both models are identical. Putting the special binding site [with energy $(-\varepsilon)$] on the first or on the second site introduces the asymmetry in the system.  For the channel in the first model the particle current is equal to\cite{AR07,kolomeisky07}
\begin{equation}
J_{1}=\frac{k_{on}(c_{1}-c_{2})}{\left[1 + \frac{k_{on} c_{1}}{k_{off}} + x^{-\theta}\left( 1+ \frac{k_{off}}{\alpha}+ \frac{k_{on} c_{2}}{k_{off}}+ \frac{k_{on} c_{2}}{\alpha} \right) + x^{1-\theta}\left( \frac{k_{on} c_{2}}{k_{off}}+\frac{k_{on} c_{1}}{\alpha}\right)+x \frac{k_{on} c_{1}}{k_{off}} \right]},
\end{equation}
where we assumed that transition rates inside the channels are the same, $u_{1}=w_{2}=\alpha$. For the model with the interaction in the second binding site one finds
\begin{equation}
J_{2}=\frac{k_{on}(c_{1}-c_{2})}{\left[1 + \frac{k_{on} c_{2}}{k_{off}} + x^{-\theta}\left( 1+ \frac{k_{off}}{\alpha}+ \frac{k_{on} c_{1}}{k_{off}}+ \frac{k_{on} c_{1}}{\alpha} \right) + x^{1-\theta}\left( \frac{k_{on} c_{1}}{k_{off}}+\frac{k_{on} c_{2}}{\alpha}\right)+x \frac{k_{on} c_{2}}{k_{off}} \right]}.
\end{equation}

The explicit expressions for dispersions $D_{1}$ and $D_{2}$ for both models can be written,\cite{AR07,derrida83} however they are quite bulky and we will not present them here. To measure the effect of the asymmetry on dynamic properties we plot the ratio of currents, $J_{1}/J_{2}$, and the ratio of dispersions, $D_{1}/D_{2}$, for both models in Fig. 6. For all quantities deviations from unity indicate that the asymmetry is important for  particle currents and for fluctuations in the number of translocated molecules. For all concentrations outside of the membrane channel the repulsive interaction ($\varepsilon <0$) on the first binding site leads to larger particle current and dispersion, although fluctuations are affected less strongly, in comparison with the situation when the repulsive interaction is on the second binding site. However, putting the attractive site first ($\varepsilon >0$) have an opposite effect: the particle currents and dispersions are generally lower for the first model. 

To understand these phenomena let us consider channels with repulsive interactions. In the first model, with the repulsion on the first binding site, the particle spends more time in the second site  because this position is energetically  more favorable for the particle. Then the molecule is closer  to the right chamber, and it might easily exit out. This  leads to  larger fluxes and dispersions. In the second model, with the repulsive interaction on the second site, the particle stays longer in the first binding site, which is further from the exit, producing smaller currents and fluctuations. These arguments suggest that in the transport of molecules across the membrane channels it is possible to control  molecular fluxes without influencing much the fluctuations by putting the binding sites at the proper positions. It  also agrees with the idea of the optimal inter-channel potential developed in continuum models of channel-facilitated transport.\cite{bezrukov07}

\section{Summary and Conclusions}

We presented a theoretical investigation of the transport of molecules through membrane pores by analyzing discrete-state stochastic models that allow one to calculate explicitly dynamic properties of the system. It was shown that interaction potentials between the molecule and the channel strongly affect translocation dynamics. For small concentrations outside of the membrane the attractive binding sites produce largest particle currents and dispersions, while for large concentrations the repulsion leads to large fluxes and fluctuations. For locally asymmetric potentials (with the interaction-distribution coefficient $\theta$ not equal to 0.5) maxima in the particle currents and dispersions are achieved for different interaction strengths, and for locally symmetric potentials ($\theta=0.5$) the largest molecular fluxes and fluctuations are taking place for the same interactions. We  found conditions for the most optimal transport across the membrane pores, when the largest possible flux is accompanied by the smallest possible fluctuations. It was shown that the most optimal interaction strength is independent of the local asymmetry in the potential, although it generally does not coincide with the positions of maximal fluxes or maximal dispersions. We also investigated the effect of the global asymmetry on translocation dynamics, and it was argued that  the location of binding sites with different interaction strengths  strongly affects the molecular transport across the channels. These phenomena are explained by using the details of dynamics of particle entering and leaving the binding sites. Our theoretical analysis suggests a possible mechanism of selectivity and efficiency of membrane channels: tuning the interaction potential by changing the interactions and asymmetry in the potential (both local and global) it is possible to control the translocation dynamics.    

It is important to note that our theoretical approach is based on several oversimplified assumptions. Specifically, interactions between the molecules and three-dimensional nature of membrane channels and interaction potentials are neglected. It will be important to compare our theoretical predictions with more realistic theoretical models and with extended molecular dynamics computer simulations. However, the most important test  of our theoretical approach should come from the experiments that will simultaneously measure molecular fluxes and fluctuations. We believe that a combination of analytical, computation and experimental methods will help to uncover the mechanisms of translocation across membrane channels.

\section*{Acknowledgments}

The authors would like to acknowledge the support from the Welch Foundation (Grant No. C-1559), the U.S. National Science Foundation (Grants No. CHE-0237105 and ECCS-0708765).

\newpage

\noindent {\bf Figure Captions:} \\
\vspace{5mm}

\noindent Fig. 1. General schematic view for  discrete-state stochastic models of channel-facilitated transport. A cylindrical membrane divides the system into three parts: the left chamber with particle concentration $c_{1}$, the right chamber with particle concentration $c_{2}$, and the pore which can be occupied by a single particle. Open circles correspond to the binding sites in the channel. At site $j$ the particle jump forward and backward with rates $u_{j}$ and $w_{j}$, respectively. The filled circle describes the position currently occupied by the particle.

\vspace{5mm}

\noindent Fig. 2 Potential of mean forces  for the channel-facilitated membrane transport. The free energy at the entrance (site 0) is equal to zero. Sites 1 and 2 are attractive, while the site 3 is repulsive. The site $N+1$ corresponds to the right chamber.

\vspace{5mm}

\noindent Fig. 3. Relative molecular fluxes and relative dispersions as a function of the interaction strength for the model with $N=1$ binding site for different concentrations and for different interaction-distribution factors. The transitions rates, $k_{on}=15$ $\mu$M$^{-1}$s$^{-1}$ and $k_{off}=500$ s$^{-1}$, are taken from Ref. \cite{schwarz03}  For all calculations $c_{2}=0$ is assumed. a) $c_{1}=10$ $\mu$M; b)  $c_{1}=500$ $\mu$M.

\vspace{5mm}

\noindent Fig. 4. Interactions producing maximal current and dispersions as a function of a) the external molecular concentration $c_{1}$; and b) the interaction-distribution parameter $\theta$. The transitions rates, $k_{on}=15$ $\mu$M$^{-1}$s$^{-1}$ and $k_{off}=500$ s$^{-1}$, are taken from Ref. \cite{schwarz03}  For all calculations $c_{2}=0$ is assumed.

\vspace{5mm}

\noindent Fig. 5. The ratio of relative dispersion over the relative current as a function of the interaction strength for the model with $N=1$ binding site for different concentrations. The transitions rates, $k_{on}=15$ $\mu$M$^{-1}$s$^{-1}$ and $k_{off}=500$ s$^{-1}$, are taken from Ref. \cite{schwarz03}  For all calculations $c_{2}=0$ is assumed.

\vspace{5mm}

\noindent Fig. 6. The ratio of current and dispersions as a function of the interaction strength for two models with $N=2$ binding sites. The transitions rates, $k_{on}=15$ $\mu$M$^{-1}$s$^{-1}$ and $k_{off}=500$ s$^{-1}$, are taken from Ref. \cite{schwarz03}  For all calculations $c_{2}=0$, $\theta=0.5$ and $\alpha=k_{off}$ are assumed.

\newpage

\begin{figure}[ht]
\begin{center}
\unitlength 1in
\begin{picture}(3.0,4.0)
  \resizebox{3.375in}{3.375in}{\includegraphics{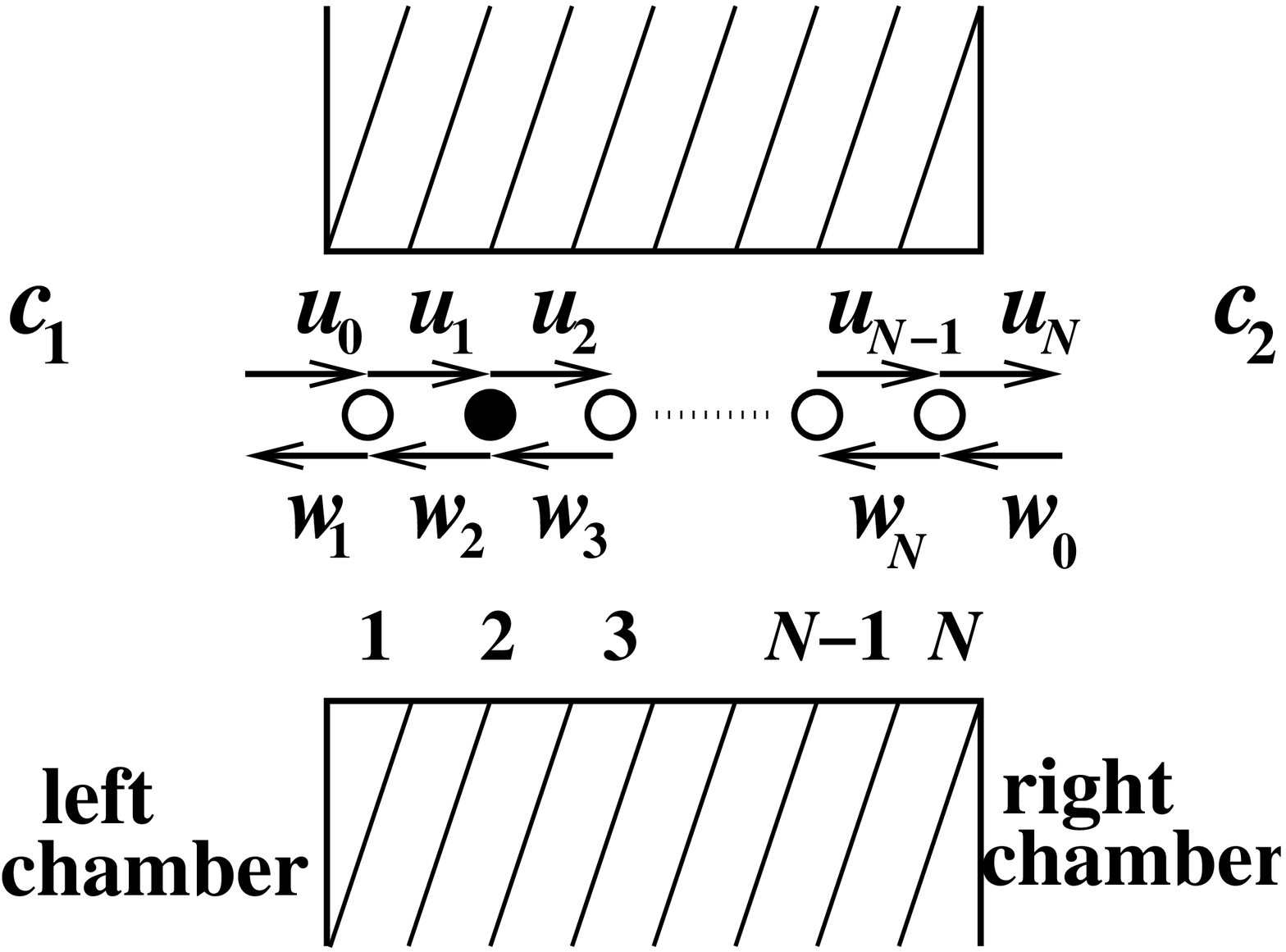}}
\end{picture}
\vskip 1in
 \begin{Large} Figure 1. Kolomeisky and Kotsev \end{Large}
\end{center}
\end{figure}

\newpage

\begin{figure}[ht]
\begin{center}
\unitlength 1in
\begin{picture}(3.0,4.0)
  \resizebox{3.375in}{3.375in}{\includegraphics{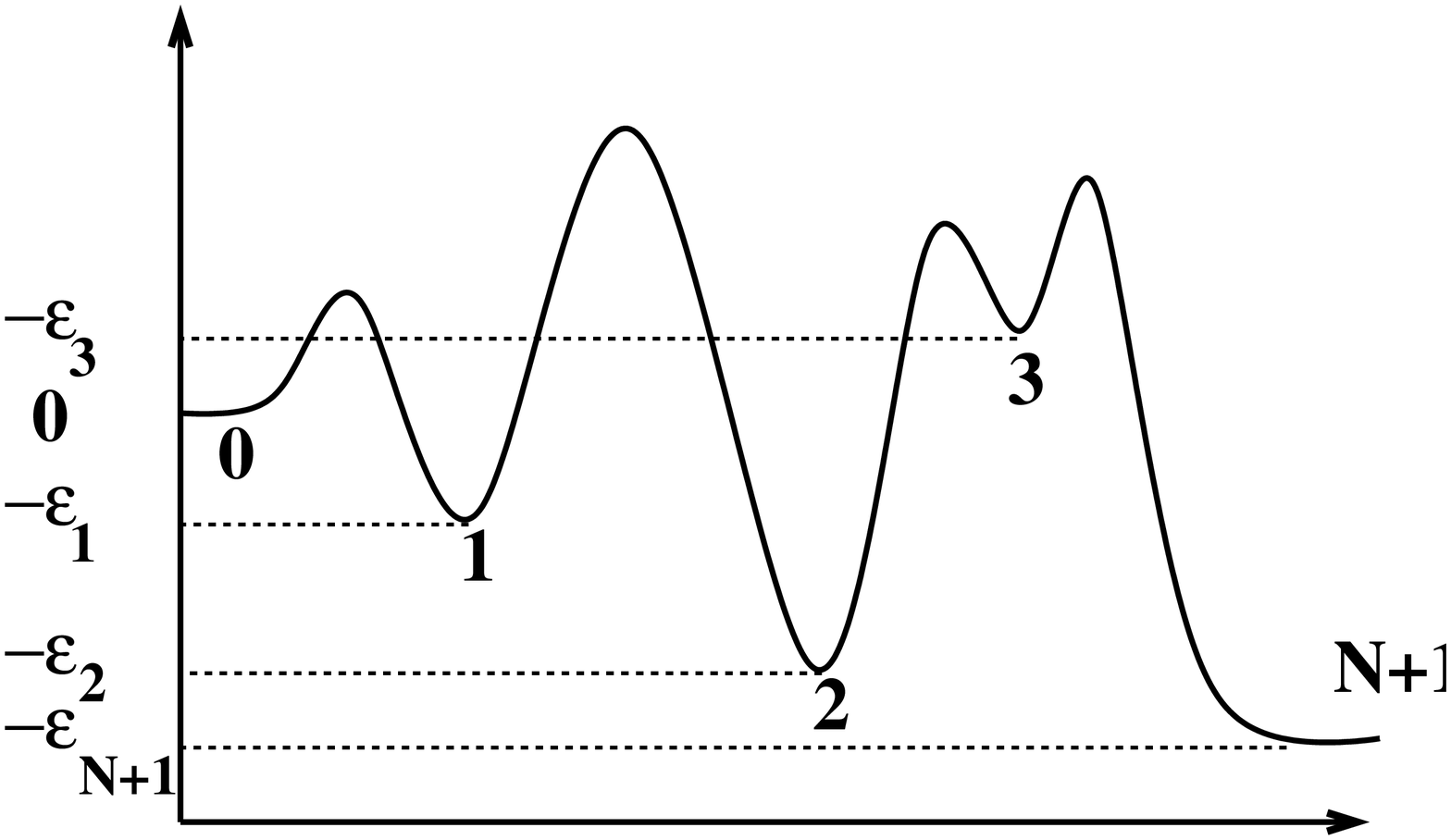}}
\end{picture}
\vskip 1in
 \begin{Large} Figure 2.  Kolomeisky and Kotsev \end{Large}
\end{center}
\end{figure}

\newpage

\begin{figure}[ht]
\begin{center}
\unitlength 1in
\begin{picture}(3.0,4.0)
  \resizebox{3.375in}{3.375in}{\includegraphics{Fig3a.eps}}
\end{picture}
\vskip 1in
 \begin{Large} Figure 3a.  Kolomeisky and Kotsev \end{Large}
\end{center}
\end{figure}

\newpage

\begin{figure}[ht]
\begin{center}
\unitlength 1in
\begin{picture}(3.0,4.0)
  \resizebox{3.375in}{3.375in}{\includegraphics{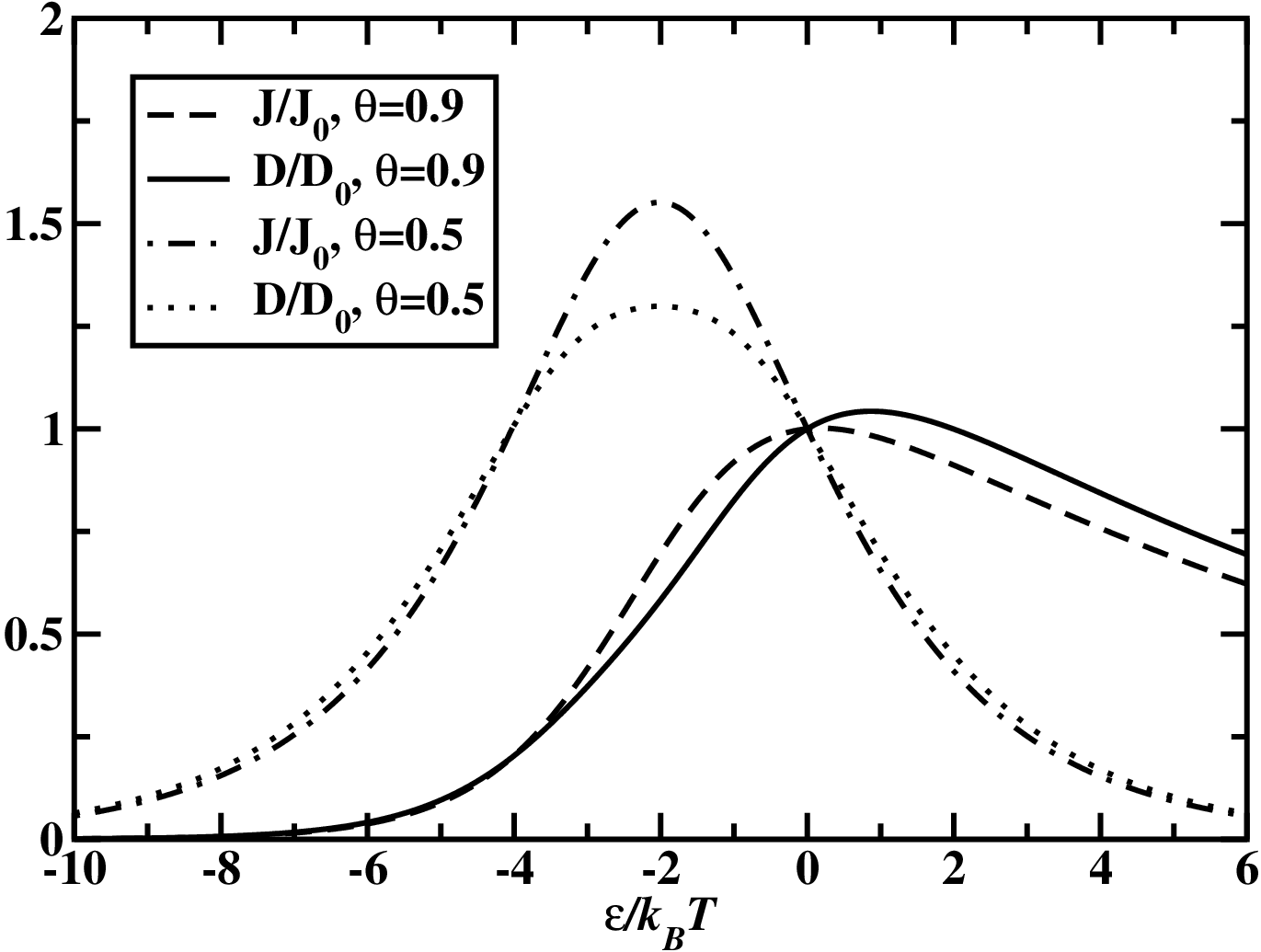}}
\end{picture}
\vskip 1in
 \begin{Large} Figure 3b.  Kolomeisky and Kotsev \end{Large}
\end{center}
\end{figure}

\newpage

\begin{figure}[ht]
\begin{center}
\unitlength 1in
\begin{picture}(3.0,4.0)
  \resizebox{3.375in}{3.375in}{\includegraphics{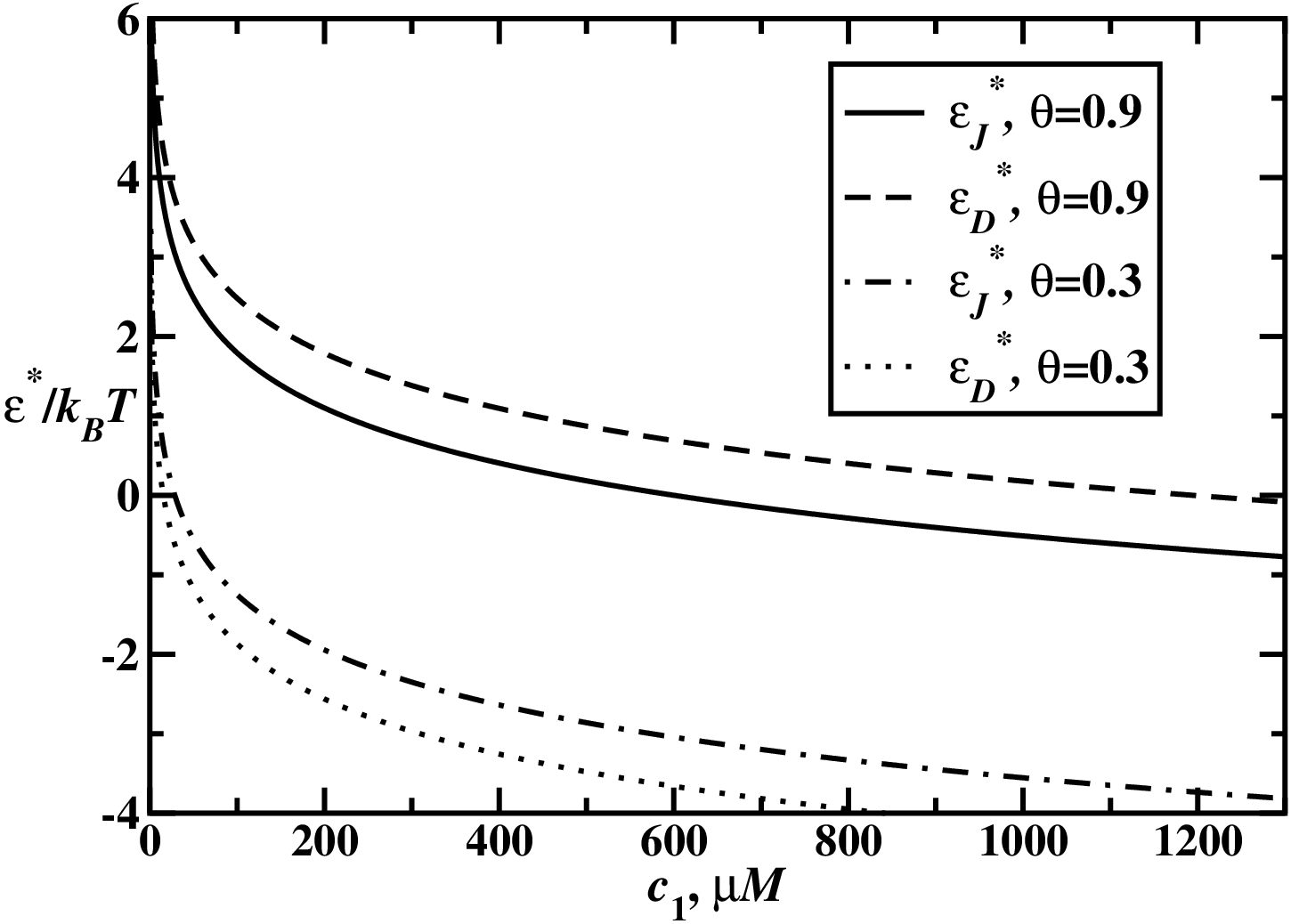}}
\end{picture}
\vskip 1in
 \begin{Large} Figure 4a.  Kolomeisky and Kotsev \end{Large}
\end{center}
\end{figure}

\newpage

\begin{figure}[ht]
\begin{center}
\unitlength 1in
\begin{picture}(3.0,4.0)
  \resizebox{3.375in}{3.375in}{\includegraphics{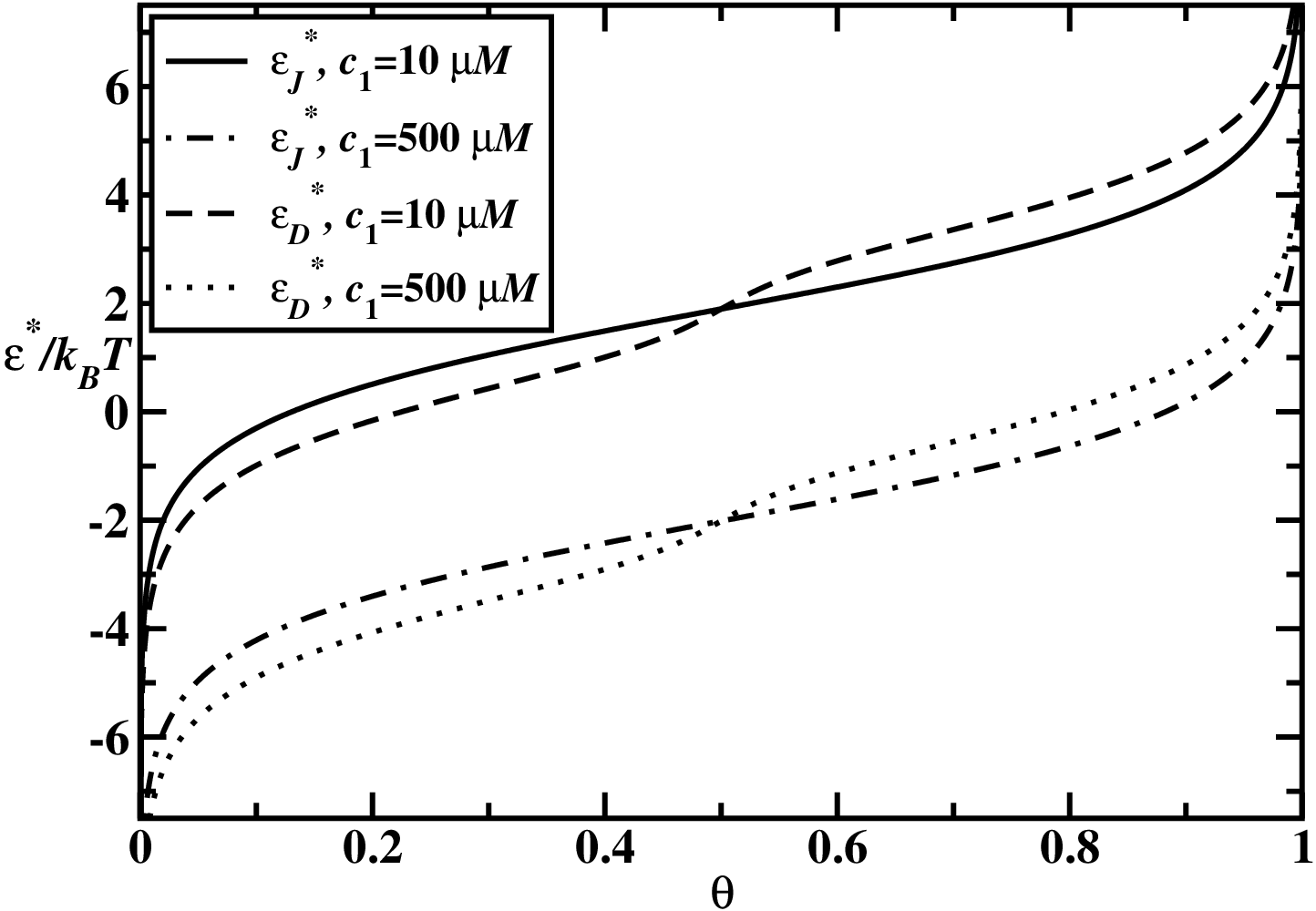}}
\end{picture}
\vskip 1in
 \begin{Large} Figure 4b.  Kolomeisky  and Kotsev \end{Large}
\end{center}
\end{figure}

\newpage

\begin{figure}[ht]
\begin{center}
\unitlength 1in
\begin{picture}(3.0,4.0)
  \resizebox{3.375in}{3.375in}{\includegraphics{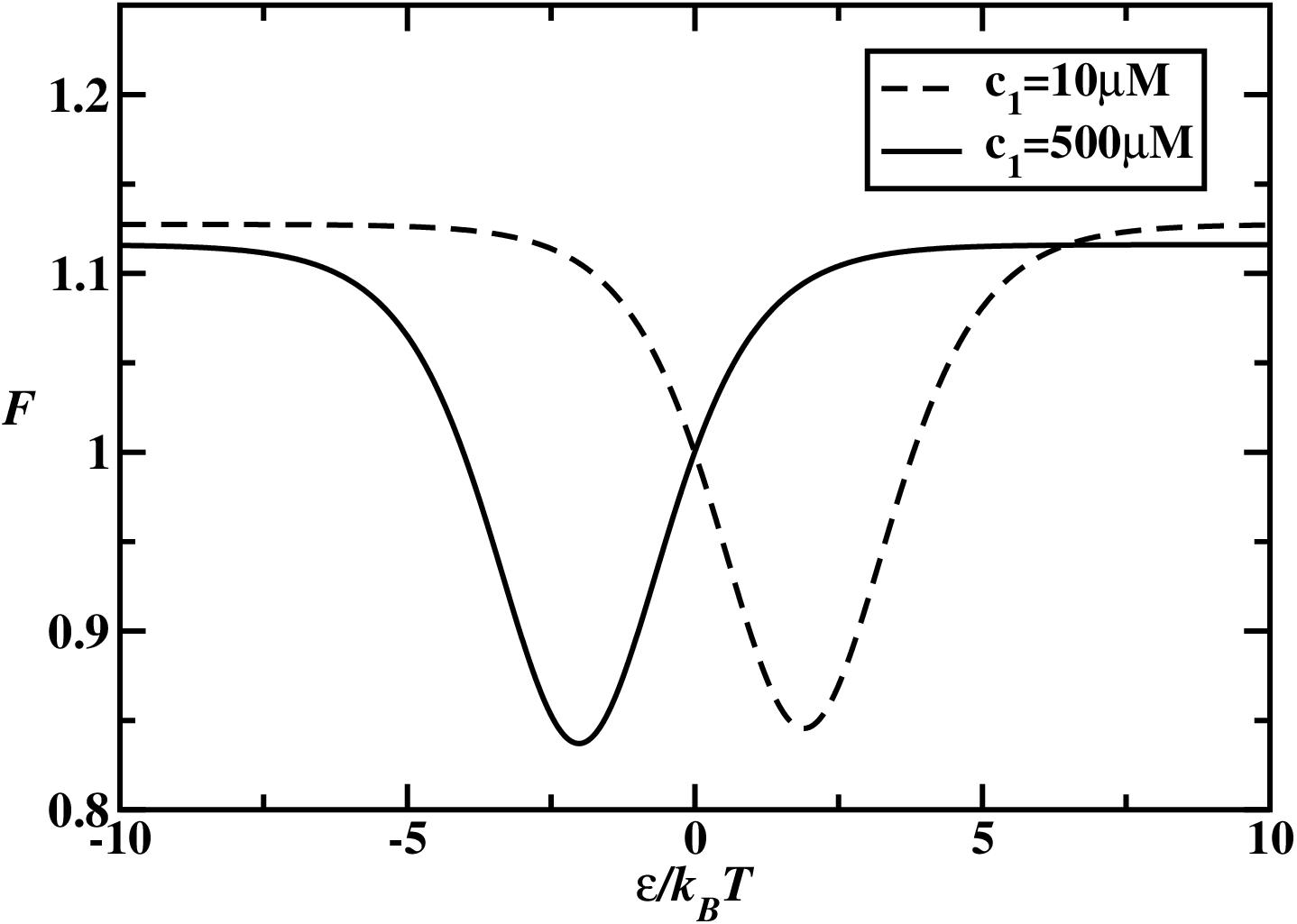}}
\end{picture}
\vskip 1in
 \begin{Large} Figure 5.  Kolomeisky and Kotsev \end{Large}
\end{center}
\end{figure}

\newpage

\begin{figure}[ht]
\begin{center}
\unitlength 1in
\begin{picture}(3.0,4.0)
  \resizebox{3.375in}{3.375in}{\includegraphics{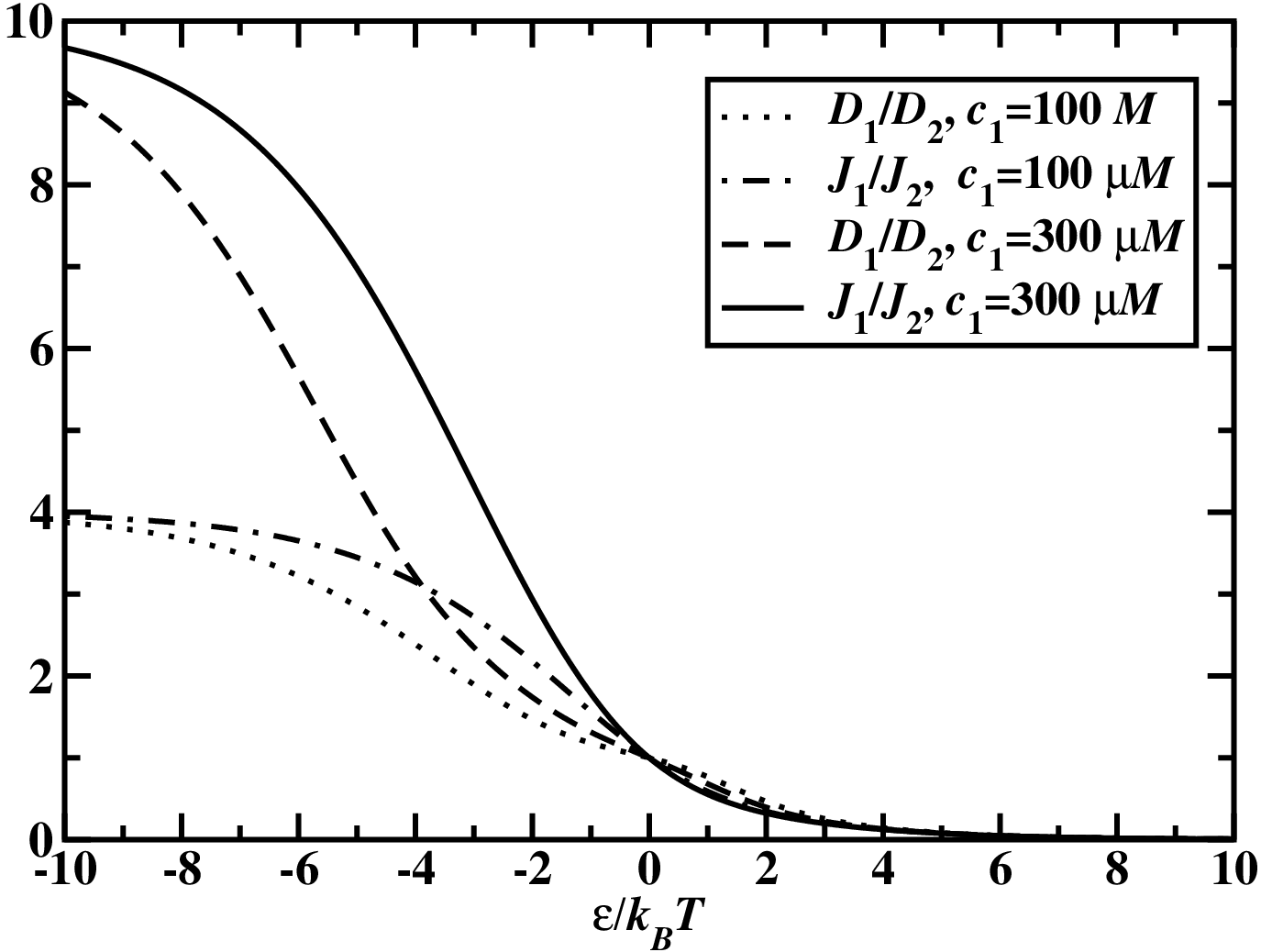}}
\end{picture}
\vskip 1in
 \begin{Large} Figure 6.  Kolomeisky and Kotsev \end{Large}
\end{center}
\end{figure}

\end{document}